\documentclass[12pt,reqno]{amsart}
\usepackage{amssymb}
\usepackage{mathrsfs}
\usepackage{dsfont}
\usepackage{amsmath,amsthm}
\usepackage{amsfonts}
\usepackage{enumerate, xspace}
\usepackage{a4wide}
\usepackage{graphicx}
\RequirePackage{hyperref}

\theoremstyle{definition}
\newtheorem{thm}{\bf Theorem}[section]
\newtheorem{prop}[thm]{\bf Proposition}
\newtheorem{defin}[thm]{\bf Definition}
\newtheorem{lemma}[thm]{\bf Lemma}
\theoremstyle{remark}
\newtheorem{rem}[thm]{\bf Remark}
\numberwithin{equation}{section}

\newcommand{\sumtwo}[2]{\sum_{\substack{#1 \\ #2}}} 

\newcommand{\dis}{\displaystyle}
\newcommand{\eps}{\epsilon}
\newcommand{\nn}{\nonumber}
\newcommand{\ga}{\gamma}
\newcommand{\Om}{\Omega}
\newcommand{\om}{\omega}
\newcommand{\si}{\sigma}
\newcommand{\Si}{\Sigma}
\newcommand{\und}{\underline}
\newcommand{\la}{\lambda}
\newcommand{\La}{\Lambda}
\newcommand{\Ga}{\Gamma}
\usepackage{color}
\definecolor{titti}{rgb}{0.09, 0.75, 0.50}
\definecolor{darkspringgreen}{rgb}{0.09, 0.45, 0.27}
\definecolor{greyd}{cmyk}{0,0,0,0.4}
	\definecolor{applegreen}{rgb}{0.55, 0.71, 0.0}
	\definecolor{darkpastelgreen}{rgb}{0.01, 0.75, 0.24}
\definecolor{electricpurple}{rgb}{0.75,0.00,1}
\newcommand{\titti}{\textcolor{electricpurple}}
\newcommand{\MMM}{\textcolor{darkpastelgreen}}
\def \blue{\color{blue}}

\def \red {\color{red}}

\def \magent {\color{magenta}}

\newcommand{\lb}{\left(}
\newcommand{\rb}{\right)}
\newcommand{\lbr}{\left\{}
\newcommand{\rbr}{\right\}}

\newcommand{\be}[1]{\begin{equation}\label{#1}}
\newcommand{\ee}{\end{equation}}

\newcommand{\fcone}{\mathcal{Y}^\blacktriangleleft}
\newcommand{\bcone}{\mathcal{Y}^\blacktriangleright}

\begin{document}


\title [ ]{Reservoirs, Fick law and the Darken effect}

\author{A. De Masi}
\address{Universit\`a dell'Aquila, Via Vetoio 1,67100 L’Aquila, Italy}
\email{anna.demasi@univaq.it}
\author{I. Merola}
\address{Universit\`a dell'Aquila, Via Vetoio 1,67100 L’Aquila, Italy}
\email{immacolata.merola@univaq.it}
\author{E. Presutti}
\address{Gran Sasso Science Institute, Viale F. Crispi 7, 67100 L’Aquila, Italy}
\email{errico.presutti@gmail.com}
\begin{abstract}
We study the stationary measures of Ginzburg-Landau (GL) stochastic processes which
describe the magnetization flux induced by the interaction with reservoirs. To privilege simplicity to generality we restrict to quadratic hamiltonians where almost explicit formulas can be derived. We discuss the case where reservoirs are represented by boundary generators (mathematical reservoirs) and compare with more physical reservoirs made by large-infinite systems.
 We prove the validity of the Fick law away from the boundaries.
We also obtain in the context of the GL models a mathematical proof of the Darken effect which shows uphill diffusion of carbon in specimen partly doped with the addition of Si.

\hfill\break
\phantom{a}\hfill\break

\end{abstract}

\date{\today}
\maketitle

\section {Introduction}
\label{sec:1}
In the context of magnetic systems (to which we restrict in this paper)
Fick law states that the magnetic current is proportional to minus the magnetization density.  Usually the Fick law is tested in a stationary setup where a magnetic fluid is in a finite cylinder  and the magnetization at its right and left faces are kept fixed at  values different from each other.  Once stationarity is reached we see a steady current which is proportional to minus the difference of the magnetization density at the boundaries.

We restrict in this paper to $d=3$ space dimension and denote  the cylinder by $T_{\ell,\ell'}=\{r=(r_1,..,r_3): |r_1| \le \ell, |r_i|\le \ell'$ for $i=2,3 \}$.
The axis of $T_{\ell,\ell'}$ has length $2\ell$ and is directed along the $r_1$ direction, the width is $2\ell'$.  Reservoirs acting on the right and left faces fix the magnetization at two distinct values, $m_{\rm right}$ and $m_{\rm left}$. The steady current is denoted by $j$, the steady magnetic profile by $m(x)$ and the Fick law states that 
\begin{equation}
 \label{DMP0.1}
j = -D(m(r))\nabla m(r)
  \end{equation}
$D(m)$ being the diffusion matrix.

\eqref{DMP0.1} has the following scaling symmetry: for any positive $\eps$
\begin{equation}
 \label{DMP0.2}
j_\eps = -D(m_\eps(r))\nabla m_\eps(r),\quad r \in T_{\eps^{-1}\ell,\eps^{-1}\ell'}
  \end{equation}
where
\begin{equation}
 \label{DMP0.3}
j_\eps = \eps^{-1} j,\quad m_\eps(r)= m(\eps^{-1}r)
  \end{equation}
 At the boundary faces $m_\eps$ has the same values  $m_{\rm right}$ and $m_{\rm left}$  independently of $\eps$.

Fick law has been derived from stochastic lattice processes using the above scaling property: of course lattice system do not have such a symmetry which however can be recovered in the limit $\eps\to 0$.
The derivation involves weak convergence and \eqref{DMP0.1} is proved only in the bulk of $T_{\ell,\ell'}$, i.e.\ except sets of zero Lebesgue measure.

Purpose of this paper is twofold;  to discuss (i) what happens in such exceptional regions and (ii) which boundary processes could be used to fix the boundary magnetization. In particular we will discuss uphill diffusion where the magnetization flows from smaller to higher values.  We start by describing the famous Darken experiment where this effect was first discovered in solids. Then in Section  \ref{sec:3} we study a quadratic stochastic Ginzburg-Landau model and show that it  reproduces the Darken effect. A simple particle model for the Darken experiments has been considered in \cite{CDP}.
 In Section \ref{sec:5} we give a physical description of the reservoirs as made by very large-infinite systems and study their stationary states.
In  Section \ref{sec:6} we draw a few concluding remarks.

\bigskip

\section {The Darken experiment}
\label{sec:2}

In his 1948 paper, \cite{darken}, Darken gave evidence of the phenomenon of uphill diffusion in metals.   We quote from \cite{darken}:

\medskip
\noindent
{\it In order to demonstrate the existence of uphill diffusion in metals, a series of four weld-diffusion experiments was made.  In these measurements pairs of steel of virtually
the same carbon content, but differing markedly in alloy content, were welded at the end and held at $1050^0C$ for about two weeks.  Subsequent analysis  showed that carbon had diffused  so as to produce an inequality of carbon content on the two sides of the weld.}
[\dots]

\begin{figure}
\includegraphics[width=.7\textwidth]{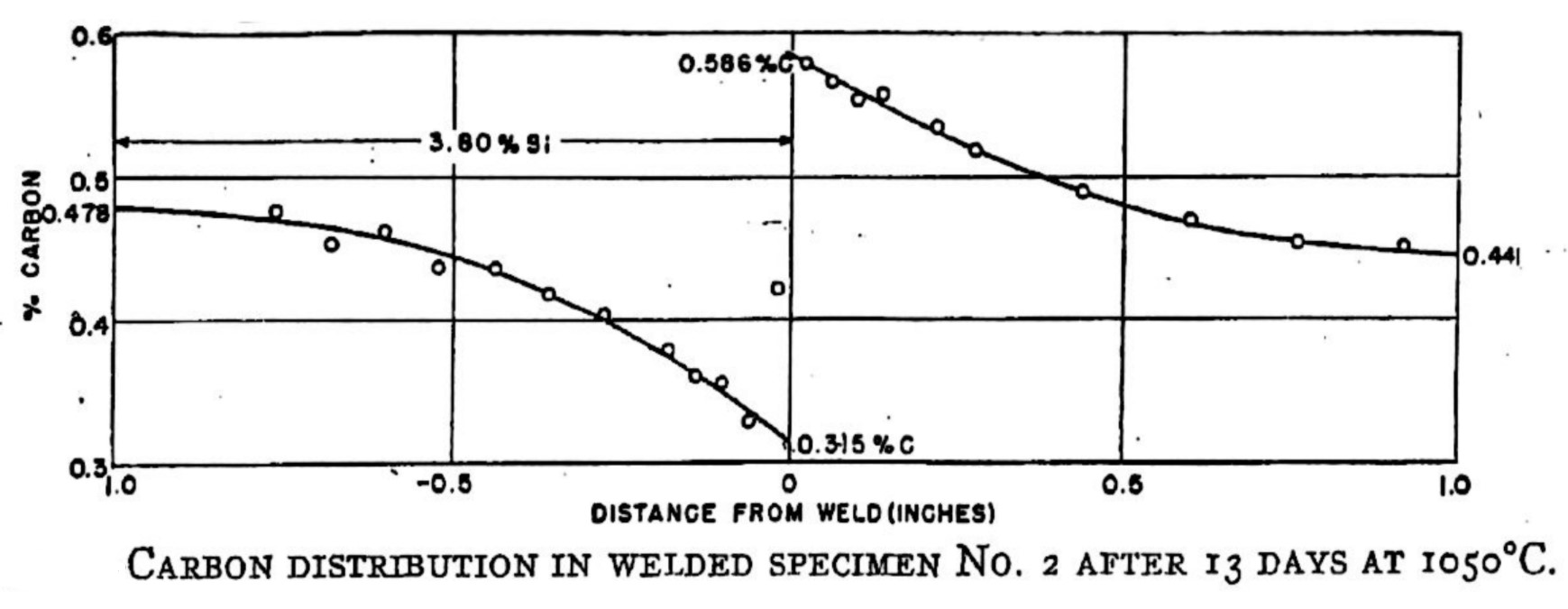}
\caption{The picture is taken from the original article of L.S. Darken, \cite{darken}}\label{fig:o-darken}
	\end{figure}

\noindent
{\it The ``uphill'' diffusion of carbon is most clear in Fig. \ref{fig:o-darken}, where it is seen that carbon diffuses from an austenite of carbon content 0.32pct to an austenite of carbon content 0.59 pct.  The difference in silicon content (3.89 and 0.05 pct respectively) is clearly responsible for the phenomenon.  [\dots] Thus silicon decreases the affinity of austenite for carbon.
}

\bigskip

\section {A model for the Darken effect}
\label{sec:3}
We will check the validity of the Darken effect in the stationary framework used to study the Fick law.
We will look at  a particular 3d stochastic Ginzburg-Landau (G-L) process where we can get almost explicit results. The stationary (non equilibrium) measure in general
G-L processes is in fact Gibbsian but with a modified hamiltonian  so that the analysis is reduced to a Gibbsian equilibrium problem.

We work in $\La_N\subset\mathbb Z^3$,  $\La_N=\La'_N\cup \La''_N$ and the two specimen in the Darken experiment are represented by the two cubes $\La'_N$ and $\La''_N$:
\begin{eqnarray}
\label{DMP3.1}
&&\La'_N:=\Big\{ (x_1,x_2,x_3)\in \mathbb Z^3:  -2N \le x_1\le -1, -N\le x_i< N, i>1\Big\},\\&&
\nn
 \La''_N:=\Big\{ (x_1,x_2,x_3)\in \mathbb Z^3:  0 \le x_1< 2N, -N \le x_i< N, i>1\Big\}
\end{eqnarray}

The phase space of the system is $\mathbb R^{\La_N}$, its elements are denoted by $\phi=\{\phi_x, x \in \La_N\}$, $\phi_x$ is regarded as a real valued magnetic moment  sitting at $x$ and directed along a fixed direction ($\phi_x$ is hereafter called the spin at $x$). The correspondence between particles and spins is the usual one: density going to 0 corresponds to spins going to $-\infty$ and density going to $+\infty$ corresponds to spins going also to $+\infty$.

As mentioned to avoid technicalities we will work with a simple hamiltonian but the result can be extended to more general GL processes.  Our  Hamiltonian is:
\begin{equation}
\label{DMP3.2}
H(\phi) := \sum_{x\in \La_N} \Big(\frac 12 {\phi_x^2} - h_x \phi_x\Big) + \frac J4\sum_{x \in \La_N} \sum_{y \in \La_N:
y\sim x } [\phi_x-\phi_y]^2,\quad J>0
\end{equation}
where $x\sim y$ means that $x$ and $y$ are nearest neighbor sites
and where
\begin{equation}
\label{DMP3.3}
h_x = \begin{cases} h&\text{if $x_1 < 0$} \\0&\text{otherwise}
\end{cases},\qquad h<0
\end{equation}
Thus positive values of the spins
are depressed in   $\La'_N$ and hence comparatively favored in $\La''_N$.  The relation with the Darken setup is the following: carbon density is replaced by magnetization density and the effect of the  Si atoms {\it which decreases the affinity of austenite for carbon} is taken into account by having added to the hamiltonian the negative magnetic field $h_x$.

We next add to the energy \eqref{DMP3.2} an interaction with the outside.  The outside is a semispace to the right, $\Om_N^{\rm right}$, and another one to the left, $\Om_N^{\rm left}$, where
   \begin{equation}
 \label{DMP5.3.0}
\Om_N^{\rm right} = \{x\in \mathbb Z^3: x_1\ge 2 N\}, \quad \Om_N^{\rm left} = \{x\in \mathbb Z^3: x_1< -2N\}
  \end{equation}
  Therefore the whole space for us is
\begin{equation}
 \label{DMP5.1.0}
\Om_N = \La_{N} \cup \Om^{\rm right}_N\cup \Om^{\rm left}_N
  \end{equation}

  \begin{figure}
\centering
\includegraphics[width=0.6\textwidth]{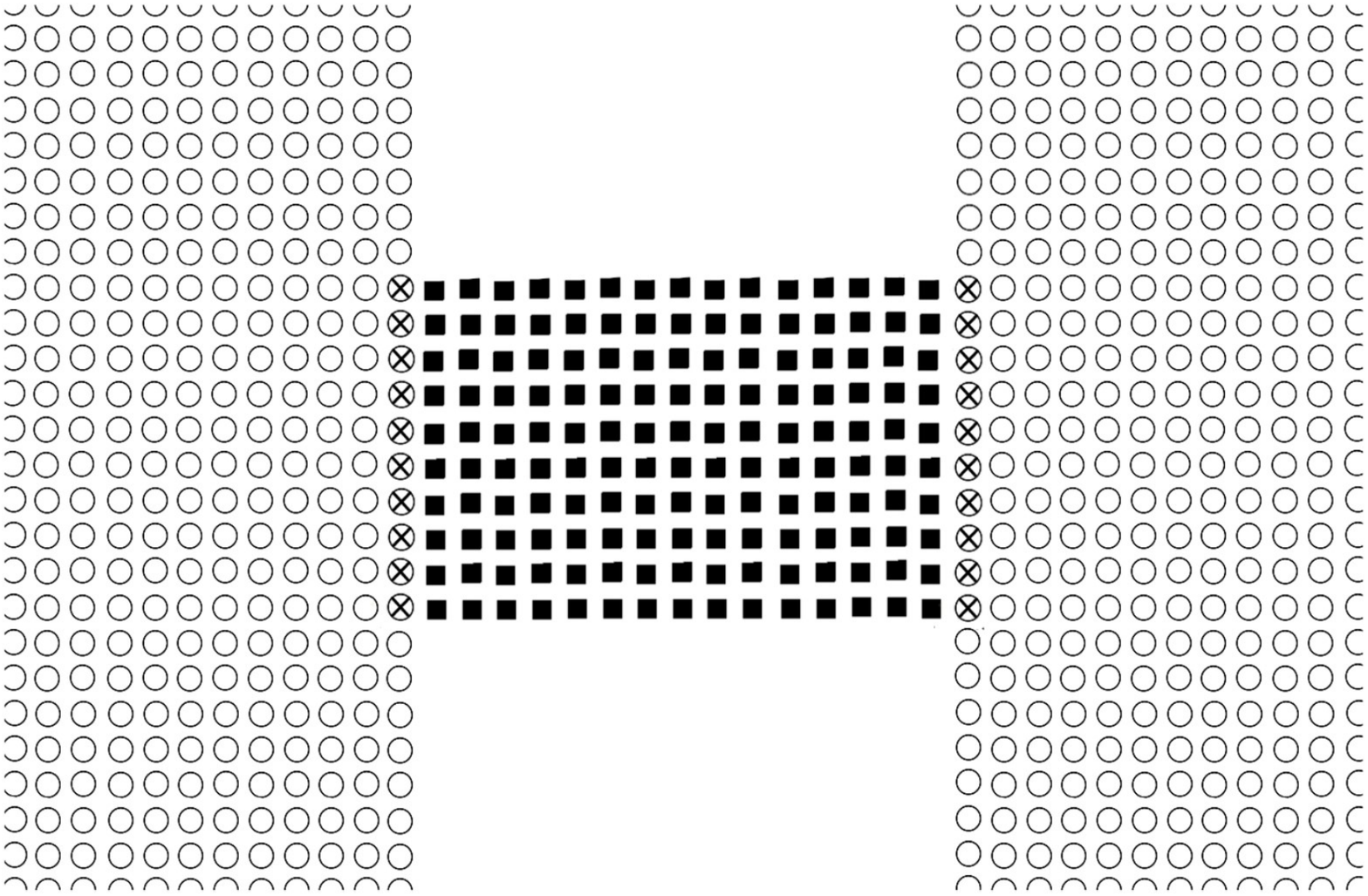}
\caption{\small The picture  shows the 2d projection of $\Om_N$. $\La_N$ is the set represented by black squares, while circles represent the points of $\Om^{\rm left}_N$ and $\Om^{\rm right}_N$, 
infinite sets,  extended respectively over all $\{x\in \mathbb Z^3: x_1<-2N\}$  and $\{x\in \mathbb Z^3: x_1\ge 2N\}$. The crossed circles represent the two sets
$\Sigma^\pm:=\{y\in \Omega^{\rm right/left}_N: \rm {dist}(y,\La_N)=1
\}$ 
defined by \eqref{DMP5.7.100}.
}\label{figOm}
	\end{figure}

The boundary condition is a configuration $\bar \phi$ on $\Om^{\rm right}_N\cup \Om^{\rm left}_N$, we take
$\bar \phi$ to be  a constant on  $\Om^{\rm right}_N$ and another constant on
$\Om^{\rm left}_N$, we choose these constants equal to $\la+h$ for $\{y_1<-2N\}$ and to
$-\la$ on  $\{y_1\ge 2N\}$, $\la \in \mathbb R$.

The new hamiltonian (modulo a constant term)  is then a function on $\phi= (\phi_y, y\in \Om_N)$
\begin{equation}
\label{DMP3.2.1.1111}
H_{\bar \phi}(\phi) =  \sum_{x\in \La_N} \Big(\frac 12 {\phi_x^2} - h_x \phi_x\Big) + \frac J4\sum_{x \in \Om_N} \sum_{y \in \Om_N:
y\sim x } [\phi_x-\phi_y]^2
\end{equation}
restricted to the set where $\phi_y=\bar \phi_y$ if $y\notin \La_N$. $H_{\bar \phi}(\phi)$ is therefore a function of $\phi_y, y \in \La_N$ and calling
\begin{equation}
\label{a3.6}
K_x = |\{ y\in \Om_N: y\sim x\}|\end{equation} we have
\begin{equation}
\label{DMP3.2.2}
H_{\bar \phi}(\phi) := \sum_{x\in \La_N} \Big(\frac 12 {\phi_x^2}[1+ K_x J] - h_x \phi_x-J\phi_x\mathbf 1_{y\notin \La_n, y\sim x}\bar \phi_y \Big) - \frac J2\sum_{x \in \La_N} \sum_{y\in \La_N:
y\sim x } \phi_x\phi_y 
\end{equation}
and we will use in the sequel the expression \eqref{DMP3.2.2}.

We will study the system in a stationary setup where
dynamics is defined in terms of a Ginzburg-Landau process. This is the Markov process with generator
   \begin{equation}
 \label{DMP3.4}
L=L^{\rm bulk}+L^{\rm left}+L^{\rm right} 
  \end{equation}
where, denoting by $x \sim y$ unordered nearest neighbor pairs in $\Om_N$,
 \begin{eqnarray}
\label{DMP3.5}
&&L^{\rm bulk} = \sum_{x \sim y \in \La_N} L_{x,y},\;L^{\rm left}= \sum_{x\in \La_N : x_1=-2N} L_{x},\;L^{\rm right}= \sum_{x\in \La_N : x_1=2N-1} L_{x}
\\&& \label{DMP3.5b}
L_{x,y}=-\Big(\frac{\partial H_{\bar \phi}}{\partial \phi_x}-\frac{\partial H_{\bar \phi}}{\partial \phi_y}\Big)\Big(\frac{\partial}{\partial \phi_x}-\frac{\partial }{\partial \phi_y}\Big)+\frac 1 \beta\Big(\frac{\partial }{\partial \phi_x}-\frac{\partial }{\partial \phi_y}\Big)^2
   \end{eqnarray}
    \begin{eqnarray}
\label{DMP3.6}
&& 
L_x=-\Big(\frac{\partial H_{\bar \phi}}{\partial \phi_x}-\la^{\rm left}\Big) \frac{\partial}{\partial \phi_x} +\frac 1 \beta \frac{\partial^2 }{\partial \phi_x^2},\quad
x : x_1=-2N
   \end{eqnarray}
       \begin{eqnarray}
\label{DMP3.7}
&& 
L_x=-\Big(\frac{\partial H_{\bar \phi}}{\partial \phi_x}-\la^{\rm right}\Big) \frac{\partial}{\partial \phi_x} +\frac 1 \beta \frac{\partial^2 }{\partial \phi_x^2},\quad
x : x_1=2N -1
   \end{eqnarray}
   ($\la^{\rm left},\la^{\rm right} \in \mathbb R$).
The generator $L_{x,y}$ describes  a process where $\phi_x+\phi_y $  is conserved and there is only an exchange of  magnetization   between the two sites. The exchange has a drift
$\frac{\partial H_{\bar \phi}}{\partial \phi_x} -\frac{\partial H_{\bar \phi}}{\partial \phi_y} $ whose effect is to decrease the energy; besides this there is a random exchange of magnetization ruled by a white noise.  In conclusion the process generated by $L^{\rm bulk}$ conserves the total magnetization and we therefore call it a Kawasaki dynamics.

The generators $L_x$ instead do not conserve the magnetization, they are Glauber-like generators.  The process generated by  $L_x$ defines a change of $\phi_x$ with a drift
toward the value $\la^{\rm right}$ if $x_1=2N-1$ and toward $\la^{\rm left}$ if $x_1=-2N$; besides this there is a white noise term. As argued in the next section these boundary processes are used in general to simulate the action of external reservoirs.

If $\la^{\rm left}=\la^{\rm right}=\la$ the whole process has a unique invariant measure which is the Gibbs measure at inverse temperature $\beta$ and hamiltonian $H_{\bar \phi}(\phi) - \sum_{x\in \La_N} \la \phi_x$.
We will instead consider the case
\begin{equation}
\label{DMP3.2.4}
\la^{\rm left} = \la,
 \; \la^{\rm right} =- \la,\quad \la>0
\end{equation}

It is  proved in \cite{DOP}, see also \cite{DIO},   that the process with the generator $L$ given in \eqref{DMP3.4} has a unique invariant measure
$d\nu_N$ which is still  a Gibbs measure at inverse temperature $\beta$ but with a modified  hamiltonian:
\begin{equation}
\label{DMP3.8}
\nu_N (d\phi)= Z^{-1}e^{-\beta [H_{\bar \phi}(\phi) - \sum_{x\in \La_N} \la_N(x) \phi_x]} \prod_x d\phi_x
\end{equation}
where $\la_N(x)=\la_N(x_1)$ and $\la_N(x_1)$ is the linear function in $[-2N-1,2N]$ with values $\la$ at $x_1=-2N-1$ and $-\la$ at $x_1=2N$:
\begin{equation}
\label{DMP3.10}
 \la_N(x_1) = \la\frac{x_1-(2N)}{(-2N-1)-(2N)}+ (-\la)\frac{x_1-(-2N-1)}
 {2N-(-2N-1)}
\end{equation}
The result in \cite{DOP} is actually valid for general stochastic Ginzburg-Landau systems with superstable hamiltonians.   Notice also that $\la_N$ is independent of the Hamiltonian.

We will use the following notation:
\begin{equation}
\label{DMP3.8.1}
H^\la_N(\phi):= H_{\bar \phi}(\phi) - \sum_{x\in \La_N} \la_N(x) \phi_x
\end{equation}
and write more explicitly
\begin{equation}
\label{DMP3.8.2}
H^\la_N(\phi) =  \frac 12 \sum_{x\in\La_N} (1+JK_x) \phi_x^2 - \frac J2 \sum_{x\in \La_N} \sum_{y\in \La_N} \mathbf 1_{x\sim y}\phi_x\phi_y - \sum_{x\in \La_N} (h_x+\la_N(x)+ J\bar \phi_y \mathbf 1_{y\notin \La_n, y\sim x})\phi_x
\end{equation}

\medskip
\subsection{The stationary magnetization profile}
\label{statmag}

The first step in deriving the Fick law is to determine the  limit  as $N\to \infty$
of the average magnetization profile $m_N(x)$:
\begin{equation}
\label{DMP3.13}
m_N(x) :=  \int \phi_x\,\,\nu_N (d\phi)
\end{equation}
%

\begin {thm}
\label{teorema3.1}
 The equation
\begin{equation}
\label{DMP3.12.5}
  \nabla_\phi H^\la_N(\phi) =0
\end{equation}
 has a unique solution which is equal to $m_N$  as defined in \eqref{DMP3.13}. Moreover

 \begin{equation}
\label{DMP3.15.0}
m_N(x)=
\alpha(x)+ \sum_{n\ge 1} \sum_{\und y =(y_1,..,y_n)} \left\{\prod_{i=0}^{n-1}\frac{K^-_{y_i}J}{1+K_{y_i}J} q(y_i,y_{i+1})\right\}\alpha(y_n)
\end{equation}
where $K_x$ is defined in \eqref{a3.6} and
\begin{subequations}\label{3.15.1}
  \begin{equation}
\label{DMP3.15.1-a}
 y_0=x,\quad  \alpha(z):=\frac{1}{1+K_{z}J}\Big(h_z+\la_N(z)+ J\mathbf 1_{y\notin \La_n, y\sim z}\bar \phi_y\Big)
\end{equation}
  \begin{equation}
\label{DMP3.15.1-b}
K^-_x = |\{ y\in \La_N: y\sim x\}|,\quad q(x,y)=
 \frac{1}{K^-_x} \mathbf 1_{y\in \La_N, y\sim x}
\end{equation}
\end{subequations}
observing that $q(x,y)$ is a transition probability.

\end{thm}

\medskip
\noindent
{\bf Proof.}
Writing explicitly \eqref{DMP3.12.5} we get the system of linear equations:
  \begin{equation}
\label{DMP3.14}
(1+ J K_x ) \psi_N(x) =   J\sum_{y\in \La_N: y\sim x}\psi_N(y) + (h_x+\la_N(x)+J \bar \phi_y \mathbf 1_{y\notin \La_n, y\sim x}),\quad x\in  \La_N
\end{equation}
We rewrite \eqref{DMP3.14} as
  \begin{equation}
\label{DMP3.15}
\psi_N(x)=\frac{K^-_xJ}{1+K_x J}\sum_{y\in  \La_N}q(x,y)\psi_N(y)  + \frac{1}{1+K_xJ}\Big(h_x+\la_N(x)+ J\mathbf 1_{y\notin \La_n, y\sim x}\bar \phi_y\Big),\quad x\in  \La_N
\end{equation}
Since
$$
\frac{K^-_xJ}{1+K_xJ} \le \frac{6J}{1+6J} <1
$$
the series obtained by iterating \eqref{DMP3.15}, which is given by the right hand side of \eqref{DMP3.15.0}, is convergent and therefore $\psi_N$ is uniquely defined.

We next Taylor expand $H^\la_N(\phi)$ around $\psi_N$:
  \begin{equation}
\label{DMP3.15.2}
H^\la_N(\phi) = H^\la_N(\psi_N) + \sum_{x\sim y, \in \La_N} a_{x,y} \tilde \phi_x \tilde \phi_y =: Q(\tilde \phi),\quad \tilde \phi_x= \phi_x - \psi_N(x)
\end{equation}
where the coefficients of the quadratic form $Q(\tilde \phi)$ are the same as in \eqref{DMP3.2} so that
$Q(\tilde \phi)$ is positive definite.
Then for any $x \in \La_N$
$$
\int e^{-\beta  Q(\tilde\phi)} \tilde \phi_x \,d\phi=0
$$
because $ Q(\tilde\phi)$ is symmetric in the exchange $\phi\to -\phi$ while $ \tilde \phi_x$ is antisymmetric. Hence $m_N(x)=\psi_N(x)$ and this  concludes the proof of \eqref{DMP3.15.0}.  \qed

\medskip
Calling $<\psi,\varphi>=\sum_{x\in\La_N}\psi_x\varphi_x$ the scalar product, we have
 \begin{eqnarray}
\label{DMP3.12.7}
&& \nu_N(d\phi) = Z_N^{-1}\exp\{-\beta\langle \tilde\phi, (D_N-A_N) \tilde\phi \rangle\} d\phi \qquad \tilde\phi = \phi-m_N\
\\&& Z_N = \Big( \frac{2\pi}{2\beta}\Big)^{|\La_N|/2} (\text{det}(D_N-A_N) )^{-1/2}\nn
	\end{eqnarray}
with $D_N$ the diagonal matrix with elements
 \begin{eqnarray}
\label{DMP3.13.2.3}
&&D_N(x,x)=(1+JK_x),\quad K_x:= \big|\{y\in \Om_N: y\sim x\}\big|
\end{eqnarray}
and $A_N$ the off diagonal matrix with elements
\begin{eqnarray}
\label{DMP3.13.2.4}
A_N(x,y) = J\mathbf 1_{y\sim x}\mathbf 1_{x,y\in\La_N}
\end{eqnarray}
Therefore the covariance matrix
 $C_N=\{c_N(x,y)\}$
  \begin{eqnarray}
\label{DMP3.12.8}
&& c_N(x,y) = \int \nu_N(d\phi) \tilde \phi(x)\tilde \phi(y)
\end{eqnarray}
is given by (see \cite{Grimmett}, Theorem 4.9.5)
 \begin{eqnarray}
\label{DMP3.12.9}
&& C_N =(2\beta[D_N-A_N])^{-1}
\end{eqnarray}

\medskip

\begin{thm}
\label{thm3.0}
For any $n\ge 1$ there are $d_n$ and $\kappa$ so that for any $x\in \La_N$
\begin{equation}
\label{DMP3.13.2.1a}
\nu_N \Big[|m_N(x)-s_N(x)| \ge \eps \Big] \le (N\eps)^{-2n}d_n \kappa^n
\end{equation}
where
\begin{equation}
\label{DMP3.13.2.1n}
s_N(x)= \frac{1}{2N} \sum_{y:y_1=x_1}\phi_y
\end{equation}

\end{thm}

\medskip
\noindent{\bf Proof.} We use
 Chebishev inequality to state that for any $n\ge 1$ there is $d_n$ so that for any $x\in \La_N$
\begin{equation}
\label{DMP3.13.2.1}
\nu_N \Big[|m_N(x)-s_N(x)| \ge \eps \Big] \le (N\eps)^{-2n}d_n \kappa^n,\quad \kappa= \sup_{x,N}\sum_y c_N(x,y)
\end{equation}
\eqref{DMP3.13.2.1a}follows from \eqref{DMP3.13.2.1} once we prove that $\kappa$ is finite.

The covariance $C_N = [2\beta(D_N-A_N)]^{-1}$  can be written as
\begin{eqnarray}
\label{DMP3.13.2.5}
C_N =(2\beta)^{-1}\sum_{n\ge 0} D_N^{-1}\{A_ND_N^{-1}\}^n
\end{eqnarray}
Thus observing that $K^-_x=6$ except at the boundaries $\{|x_2|=N\}\cup\{|x_3|=N\}$ where it is equal to 4 or 5 so that $\dis{\frac{JK^-_x}{1+JK_x}\le \frac{6J}{1+6J}}$, we get that $ \kappa$ in \eqref{DMP3.13.2.1} is $\kappa=(2\beta)^{-1} $.\qed

\bigskip
We next study the macroscopic limit of $m_N$.

\begin{lemma}
\label{lemma3.3}
 There is a constant  $c$ so that the following holds.   Let
 \begin{equation}
	\label{DMP3.18.11}
u_N(x)=\frac{ h_x + \la_N(x)}{1+6 J}\sum_{n= 0}^{\log N}(\frac{6J}{1+6J})^n \sum_{y_1,..,y_n\in \La_N} p(x,y_1)\cdots p(y_{n-1},y_n)
		\end{equation}
		where 
\begin{equation}
	\label{DMP?.0.100.4.3}
 p(x,y)= \frac{1}{6} \mathbf 1_{y\sim x}
		\end{equation}
Then for all
 $x\in \La_N$ such that  dist$\big(x,\partial \La_N\cup \{x_1=0\}\big)> \log N$.
\begin{equation}
	\label{mu3.30}
|m_N(x) - u_N(x)|\le c
\Big( \om^{\log N} + \frac{\la}{N}\log N\Big)
		\end{equation}
where $\dis{\om = \frac{6J}{1+6J}}$.

\end{lemma}

\medskip
\noindent{\bf Proof.} Recalling that $m_N(x)$ is given in  \eqref{DMP3.15.0} we first observe that if $y\in\La_N$ is such that dist$(y,\partial \La_N\cup \{x_1=0\})\ge\log N$ then $K_y=6$,  $q(y,y')=p(y,y')$ for all $y'$ and moreover $\dis{\alpha (y)=\frac{h_y+\la_N(y)}{1+6J}}$. Thus from \eqref{DMP3.15.0} we get
	\begin{eqnarray}
	\nn
m_N(x)-u_N(x)&=& \sum_{n= 0}^{\log N}(\frac{6J}{1+6J})^n
\sum_{y_1,..,y_n\in \La_N} p(x,y_1)\cdots p(y_{n-1},y_n)
\big[\alpha(y_n)-\frac{ h_x + \la_N(x)}{1+6J}\big]
\\&&+ \sum_{n\ge \log N} \sum_{\und y =(y_1,..,y_n)} \left\{\prod_{i=0}^{n-1}\frac{K^-_{y_i}J}{1+K_{y_i}J} q(y_i,y_{i+1})\right\}\alpha(y_n)\label{a3.31}
	\end{eqnarray}
Since $K_y\le 6$ for all $y$, the second term on the right hand side of \eqref{a3.31} is bounded by $c\om^{\log N}$.    $y_n$ in the first term is such that $|y_n- x|\le c\log N$ and dist$(y_n,\partial \La_N\cup \{x_1=0\})>\log N$, thus $h_{y_n}=h_x$ and
$\dis{|\la_N(y_n)-\la_N(x)| \le c \log N\frac{\la}{N}}$ which implies that the first term on the right hand side of \eqref{a3.31} is bounded by the second expression in \eqref{mu3.30}. \qed

\medskip
Next theorem is a direct consequence of Lemma \ref{lemma3.3}.

\begin{thm}
\label{thm3.0.1}
 Let $x_N$ be the integer part of  $N  r$ with  $r\in \{  |r_i|<2\}\setminus\{r_1=0\}$
 then
\begin{equation}
	\label{DMP3.18}
\lim_{N\to \infty}m_N(x_N)= :m(r) =  \la(r) + h \mathbf 1_{r_1<0}
		\end{equation}
where
		\begin{equation}
\label{DMP3.19}
\lim_{N\to \infty} \la_N(x_N) = : \la(r)= -\frac \la 2 r_1
		\end{equation}

\end{thm}

\medskip
\noindent{\bf Proof.} From the definition \eqref{DMP3.10} we get  \eqref{DMP3.19}. Observing that, for $N$ large enough,  $x_N$ satisfies the hypothesis of Lemma \ref{lemma3.3} we can use \eqref{mu3.30} and observe that
	\begin{equation*}
\Big |u_N(x_N)-\{ \frac{ h_{x_N} + \la_N(x_N)}{(1+6 J)}\}\sum_{n\ge 0}(\frac{6J}{1+6J})^n \sum_{y\in \mathbb Z^3}p^n(x_N,y)\Big |\le c\om^{\log N}
	\end{equation*}
$p^n$ being the $n$-step transition probability. The series in the above equation converges to $1+6J$, thus from \eqref{mu3.30}  we get
	\begin{equation}
	\label{mn}
	\lim_{N\to\infty} \big| m_N(x_N)- h_{x_N} - \la_N(x_N)\big |=0
	\end{equation}
	concluding the proof of the Theorem. \qed
	\medskip
	
\begin{figure}[h]
\vskip -1cm
  \centering
\includegraphics[width=10cm]{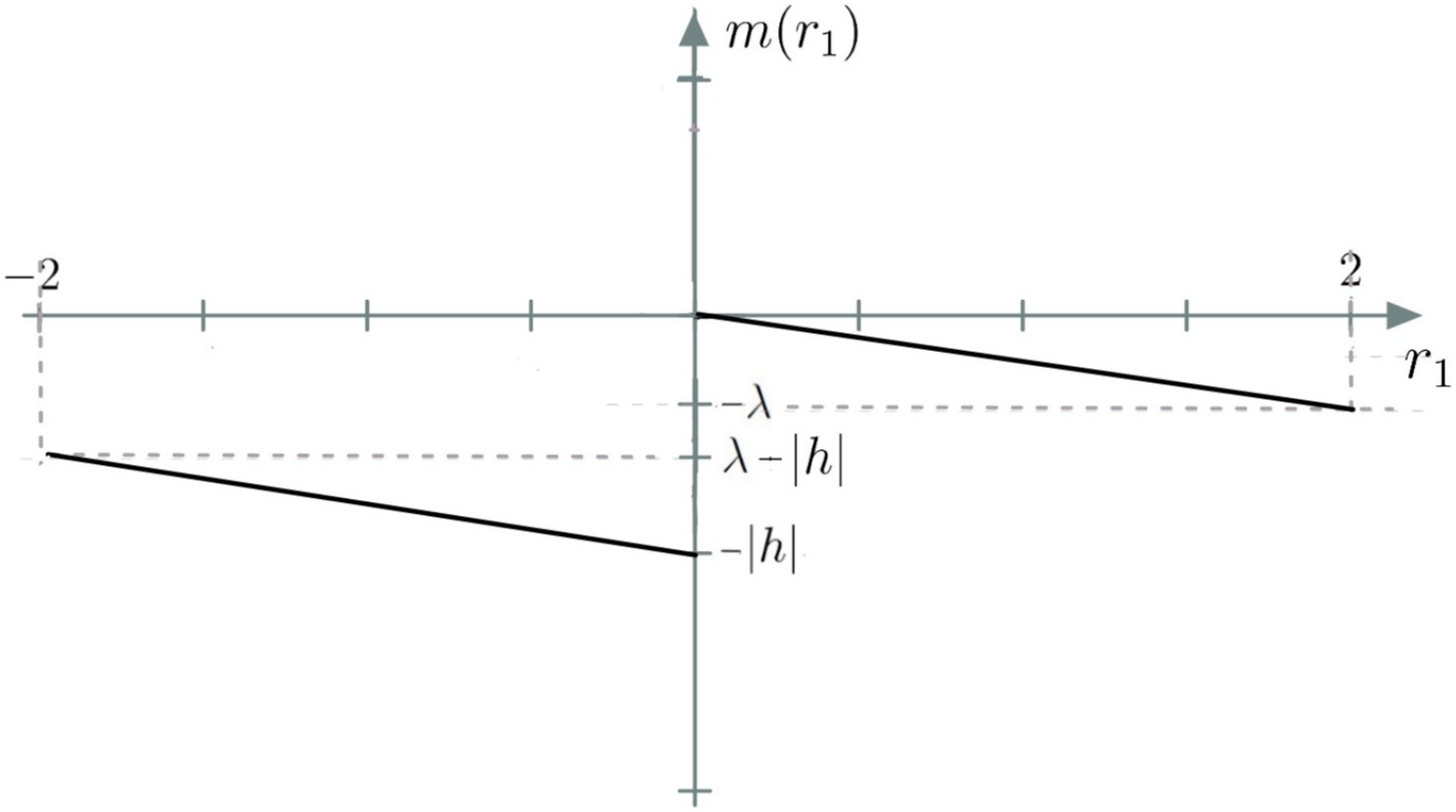}
 \vskip -1cm \caption{ }\label{m-macroprofile}
\end{figure}

The function  $m(r)$ depends only on $r_1$ and it is visualized in Fig.\ref{m-macroprofile} in the case when $\la-|h|<-\la$

\subsection{Fick law}
\label{fl}
To  check the Fick law we need to relate the limit magnetization profile to the limit current, which is computed next.

The instantaneous current $I'_{x\to y}$, $x \sim y$, through the
bond $(x,y)\subset \La_N$  is defined as
\begin{equation}
	\label{DMP3.10.0}
I'_{x\to y}=-L_{x,y}\phi_x
		\end{equation}
\begin{thm}
\label{thm3.0.0}
The stationary current through the bond $(x,y)$  is
				\begin{equation}
		\label{DMP3.18.1}
I^{(N)}_{x\to y}:= \int  I'_{x\to y} \,\, \nu_N(d\phi) =\la_N(x)-\la_N(y)
\end{equation}
Observe that $I^{(N)}_{x\to y}=0$ if $y_1=x_1$.
\end{thm}

\medskip
\noindent
{\bf Proof.} Recalling \eqref{DMP3.8} we have
 $$
   \int \frac{d}{d\phi_x}\Big(H_{\bar \phi}-\la_N(x)\phi_x\Big) \,\, \nu_N(d\phi) = - \frac 1Z   \int \frac 1\beta   \frac{d}{d\phi_x} e^{-\beta [H_{\bar \phi}-\sum_x\la_N(x)\phi_x]}\,d\phi =0 $$

 Thus
   $$
    \int d\nu_N\frac{d}{d\phi_x} H=\la_N(x)
   $$
 Recalling \eqref{DMP3.5},
   $$
    \int  L_{x,y}\phi_x \,\, \nu_N(d\phi)= - \int d\nu_N\Big(\frac{d H}{d\phi_x}-\frac {d H}{d\phi_y} \Big)= -[\la_N(x)-\la_N(y)]
   $$
  hence, recalling \eqref{DMP3.10.0}, we get \eqref{DMP3.18.1}.
  \qed

  \medskip
%
We now relate the macroscopic profile to the macroscopic stationary current. Let $r $, $|r_1|<2$, $|r_i|<1$, $i=2,3$ and call $x_N$ the integer part of $Nr$,  then by \eqref{DMP3.18.1} and \eqref{DMP3.19}
	\begin{equation}
	\label{a3.44}
I(r_1):=\lim_{N\to \infty} N I^{(N)}_{x_N\to y_N}= \frac \la 2,\qquad  y_N= x_N+e_1
		\end{equation}
$I(r_1)$ is the current in the horizontal direction, the other components of the current being equal to 0 as we observed in Theorem \ref{thm3.0.0}.

Thus by \eqref{DMP3.18} and \eqref{a3.44}
\begin{equation}
	\label{DMP3.188.2}
\frac{dm}{dr_1} = -  I(r_1)
		\end{equation}
so that the Fick law is satisfied with diffusion coefficient equal to 1.
 Calling $F_\beta$ the equilibrium free energy we are going to prove that
\begin{equation}
	\label{DMP3.188.3}
  F''_\beta(m) =1		\end{equation}
So that $\dis{\frac{dm}{dr_1} = - F''_\beta(m) I}$ in agreement with the Fick law.

Call $\pi_\beta(\zeta)$ the thermodynamic pressure when the chemical potential is $\zeta$.  Since  $\pi'_\beta(\zeta) = m_\beta(\zeta)$, the latter the equilibrium magnetization, it follows from \eqref{DMP3.18} that
\begin{equation}
	\label{DMP3.188.4}
 \pi''_\beta(\zeta)= 1  = (F''_\beta)^{-1}
		\end{equation}
(\eqref{DMP3.188.4} because pressure and free energy are Legendre
conjugate). \eqref{DMP3.188.3} then follows.   \qed

\begin{figure}[h]
\vskip-.5cm
  \centering
\includegraphics[width=5cm]{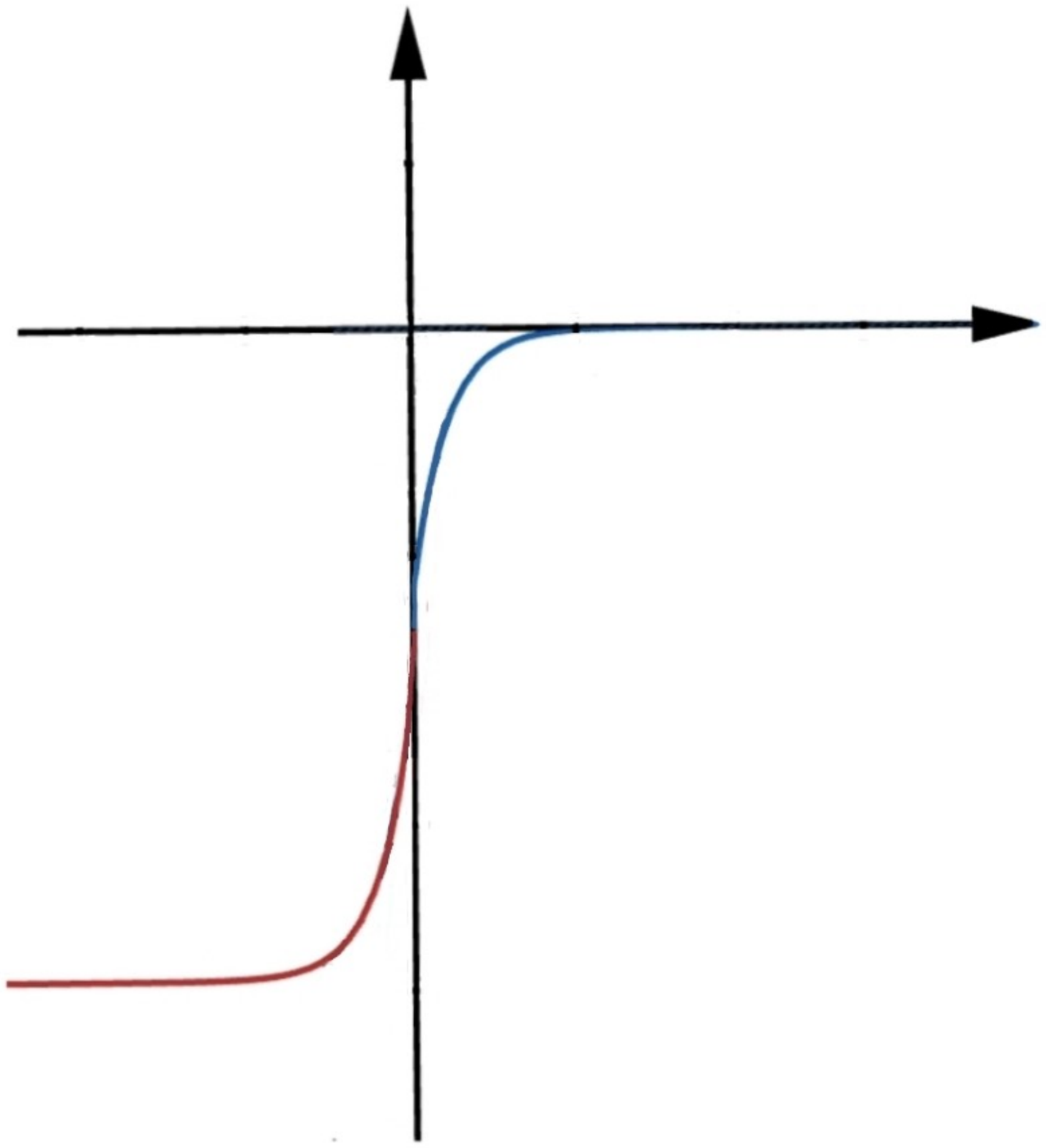}
 \vskip -1cm \caption{}\label{fig:darken}
\end{figure}

\subsection{Boundary effects}
The limit procedure used in the derivation of the limit profile $m(r)$ does not say what happens close to $|r_1| = 2$ and to $r_1=0$ where $h_x$ has a  discontinuity, questions that are answered in this subsection.
We
start with a neighbor of the vertical plane $x_1=0$ which is the most relevant in relation to the Darken effect, the result is described in Fig.\ref{fig:darken}.

\medskip

\begin{thm}
\label{teorema3.9}
Let $x\in \La_N \setminus \partial \La_N$
and $N$ so large that the distance of $x$ from $\partial \La_N$ is larger than $\log N$.
Then
	\begin{equation}
	\label{teo3.42}
\lim_{N\to\infty} m_N(x)=\bar m(x),\qquad \bar m(x) =\sum_{n\ge 0}(\frac{6J}{1+6J})^n \sum_{y\in \mathbb Z^3}p^n(x,y) \frac {h_y}{1+6J}
	\end{equation}
where, recalling \eqref{DMP?.0.100.4.3}, $p^n$ is the $n$-step transition probability.
\end{thm}

\medskip
\noindent
{\bf Proof.} We proceed as in Lemma \ref{lemma3.3}
 replacing $u_N$ by
\begin{equation}
	\label{DMP?.0.100.4.2}
 v_N(x):= \sum_{n= 0}^{\log N}(\frac{6J}{1+6J})^n \sum_{y_1,..,y_n\in \La_N}\Big(p(x,y_1)\cdots p(y_{n-1},y_n)\Big)\frac{h_{y_n} }{1+6J}
		\end{equation}
The difference from $u_N$ is that $\la_N$ is missing and instead of $h_x$ we have here $h_{y_n}$. However
$\dis{|\la_N(y)-\la_N(x) |\le c \frac{|y-x|}{N}}$,  $|y-x| \le \log N$. Moreover since
$\la_N(0)=0$, $\dis{|\la_N(x) |\le c \frac {|x|}{N}}$. Thus
	\begin{equation*}
|m_N(x)-v_N(x)|\le c\Big( \om^{ \log N} +\frac {\la \log N}{N} +\frac {|x|}N   \Big)
	\end{equation*}
Since $\dis{\lim_{N\to\infty} v_N(x)=\bar m(x)}$ we get \eqref{teo3.42}.\qed

\begin{thm}
\label{thm3.2} [The magnetization at the junction]
Let $x_1\ge 0$ then
\begin{equation}
\label{DMP3.21}
\bar m(x) =  \bar m(0)\; e^{-\ga x_1},\quad
\frac h2 < \bar m(0)<0
\end{equation}
where
 $\ga>0$ is the positive solution of
\begin{equation}
\label{DMP3.23.1}
 3= \frac {6J}{1+6J}\Big( 2  +  \cosh(\ga)\Big)
\end{equation}

For $x_1<0$
\begin{equation}
\label{DMP3.21.1.1}
\bar m(x)=  -|h| +|\bar m(0)| \, e^{-\ga (|x_1|-1)}
\end{equation}

\end{thm}

\medskip
\noindent
{\bf Proof.}  Recalling \eqref{teo3.42} and since $h_y=-|h|\mathbf 1_{y_1<0}$ we have
\begin{equation}
\label{DMP3.25}
 \bar m(x) = -\frac{|h| } {1+6J}\Big(\sum_{n\ge 0}(\frac {6J}{1+6J})^n \sum_{y_1<0}p^n(x,y)\Big)
\end{equation}
For any $n\ge 1$
$$
\sum_{y_1<0}p^n(0,y) < \frac 12
$$
 so that  $\bar m(0)< -\frac 12 |h|$.

 We observe that for any $x\in\La_N$, $\bar m(x)=\bar m(x_1)$  verifies
 	\begin{eqnarray}
\label{DMP3.22}
&& \bar m (x_1) = \frac {6J}{1+6J}\Big( \frac 23 \bar m(x_1) +  \frac 16 \bar m(x_1-1)+  \frac 16 \bar m(x_1+1)\Big) - \frac{|h| } {1+6J} \mathbf 1_{x_1<0}
	\end{eqnarray}
By the choice of $\ga$ one can check that
\begin{equation}
\label{DMP3.23}
\bar m(x_1) = \bar m(0)\, e^{-\ga x_1}, \quad x_1\ge 0
\end{equation}
 \eqref{DMP3.21.1.1} follows from  \eqref{DMP3.21} by a symmetry argument.   \qed

 \bigskip

 We conclude the section by studying the behavior of $m_N(x)$ when $x$ is close to the left boundary $x_1=-2N$, (the analysis when it is close to the right boundary is similar and omitted).

We fix $x$, $x_1\ge 0$, and we want to compute $m_N(-2N e_1 + x)$ for $x$ such that $-2N e_1 + x\in \La_N$  and away from  $\{|x_2|= N\}\cup\{ |x_3|=N\}$.  We use \eqref{DMP3.14} recalling that $\psi_N$ in that equation has been proven to be equal to $m_N$.  Calling $\xi=-2Ne_1+x$
 \begin{equation}
\label{DMP3.14.000}
(1+6J) m_N(\xi) =   J\sum_{y\in \La_N: y\sim \xi}m_N(y) + (h+\la_N(\xi)+J \bar \phi_y \mathbf 1_{y\notin \La_n, y\sim \xi})
\end{equation}
By taking the limit $N\to \infty$, $\la_N(\xi)\to \la$ and since $h+\la$ satisfies the above equation with $\la$ in place of $\la_N$, we have that
$m_N(-2Ne_1 + x) \to h+\la$. Thus the choice of the boundary conditions, $\bar \phi=h+\la$
is responsible for not having boundary layers at the right and left boundaries.

\subsection{Boundary layers and uphill diffusion}

\begin{figure}[h]
\vskip -.5cm
  \centering
\includegraphics[width=5cm]{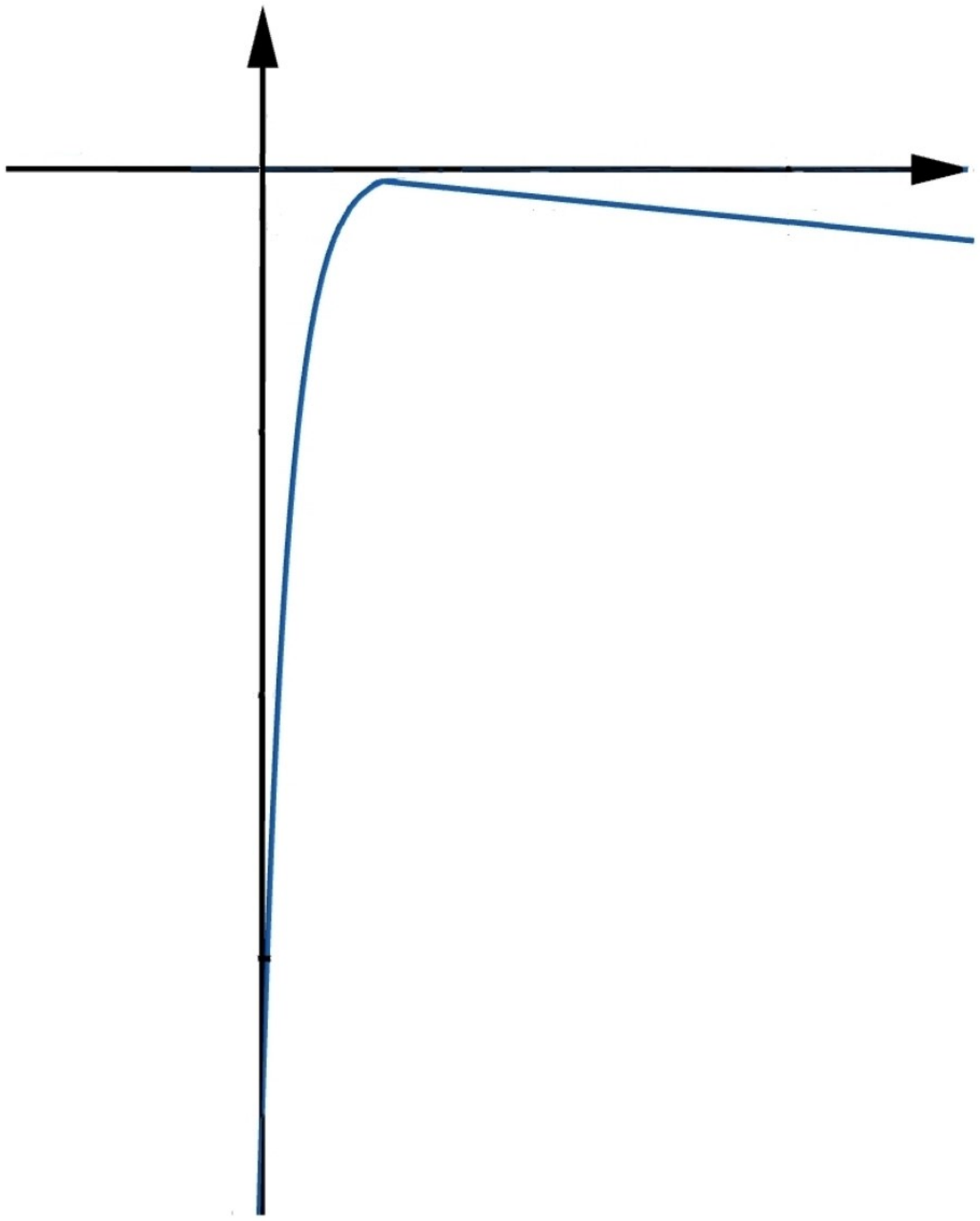}
  \vskip -.5cm\caption{}\label{fig:3-4}
\end{figure}

The analysis in this section has shown that the behavior of the magnetization at the junction, i.e.\ where $x_1$ is close to $x_1=0$ has the same features of uphill diffusion observed experimentally by Darken.   In particular if we look at the profile in the region $x_1\ge 0$, see Fig \ref{fig:3-4},
we see an increasing profile with a positive current in a region where $h=0$ and there is no external force.  All that is not against the Fick law because the phenomenon involves a region which has vanishing measure in the macroscopic limit $N\to \infty$.  When $N$ is still finite we cannot even talk about the gradient of the magnetization  because the magnetization is defined on the lattice.
%

In order to take derivatives we should have a smooth magnetizaton profile (in the limit $N\to \infty$) and this is not what happens here.  We have results in a paper in preparation by the present authors and D. Ioffe where we observe the presence of a second scale also divergent with $N$ but infinitesimal with respect to $N$ where the magnetization is not the same as that predicted by the Fick law.  This happens close to the right and left boundaries of the domain in the presence of phase transitions and when the rate of injection-removal at the boundaries is very small (infinitesimal as $N\to \infty$). The phenomenon is due to metastability effects.

\section {Physical reservoirs}
\label{sec:5}

The current flux in the previous section was determined by the action of the boundary generators $L^{\rm left}$ and $L^{\rm right}$.  They are supposed to describe the action of reservoirs which interact with the system and our next aim is to check the validity of such a statement.

The physical intuition is that reservoirs are extremely large systems with space-time correlations which decay very fast. Thus in a finite time window
 their magnetization is essentially constant  so that the system sees the reservoirs always at equilibrium. However all that may seems implausible when applied to
 stationary measures as this requires that the reservoirs should not change even if one gives and the other one receives an infinite amount of magnetization.  Indeed if the dimensions of the space is $d\le 2$ the state of the reservoirs does change while it does not in a special geometry in $d\ge 3$ as proved in this section for $d=3$.

\medskip

 The physical space   is
\begin{equation}
 \label{DMP5.1}
\Om_N = \La_{N,M} \cup \Om^{\rm right}_N\cup \Om^{\rm left}_N
  \end{equation}
where $M=N^{\alpha}$, $\alpha \in (0, \frac 12)$ (as we shall see the assumption on $M$ is essential in our analysis),
$\La_{N,M}$ is equal to
\begin{equation}
 \label{DMP5.2}
 \La_{N,M} =\{ x\in \mathbb Z^3: |x_1|< N, |x_2| \le M, |x_3| \le M\}
   \end{equation}
   \begin{equation}
 \label{DMP5.3}
\Om_N^{\rm right} = \{x\in \mathbb Z^3: x_1\ge N\}, \quad \Om_N^{\rm left} = \{x\in \mathbb Z^3: x_1\le -N\}
  \end{equation}
The phase space of our system is then $\mathbb R^{\Om_N}$, while $\mathbb R^{\Om^{\rm right}_N}$ and $\mathbb R^{\Om^{\rm left}_N}$ are the phase space of the right, respectively left reservoirs.  The formal hamiltonian $H(\phi)$ in $\Om_N$ is given by
\begin{equation}
\label{eq4.4}
H(\phi) := \sum_{x\in \Om_N} \frac 12 {\phi_x^2} + \frac J4\sum_{x \in \Om_N} \sum_{y \in \Om_N:
 |x-y|=1 } [\phi_x-\phi_y]^2,\quad J>0
\end{equation}
which differs from the Hamiltonian in
 \eqref{DMP3.2} because: (i) the sums are extended to $\Om_N$ and (ii)
$h_x\equiv 0$.  Thus our system interacts and exchanges magnetization with the reservoirs.   Dynamics is defined by the formal generator
  \begin{equation}
 \label{DMP5.3.1}
L^*= \frac 12 \sum_{x\in \Om_N}\;\;\sum_{y\in \Om_N: |x-y|=1}L_{x,y}
  \end{equation}
where $L_{x,y}$ is defined in \eqref{DMP3.5b}. Existence of dynamics is proved in \cite{CGP}
together with the existence of invariant measures.

In this section we will
compare the stationary magnetization  restricted to $\La_{N,M}$ with that of Section \ref{sec:3} with $\La_{N}$ replaced  by $\La_{N,M}$ and with zero boundary conditions. We will prove equality in the limit $N\to\infty$.

\bigskip




The following Theorem, proved in  \cite{CGP}, gives a sufficient condition for a measure to be invariant.

\begin{thm}
 \label{teorema5.1} Let  $\la^*_N(x)$ be a bounded harmonic function in $\Om_N$ (see Definition \ref{defin4.2} below) and let  $\nu^*_N$ the DLR measure at the inverse temperature $\beta$ and with formal hamiltonian      \begin{equation}
 \label{DMP5.7}
H(\phi) - \sum_x \la_N(x)\phi_x
  \end{equation}
Then $\nu^*_N$ is invariant for the process with generator $L$.
\end{thm}

\begin{defin}
\label{defin4.2}
Let $x(t)$, $t\ge 0$, be the  continuous time random walk on $ \Om_N$ with generator 
  \begin{equation}
 \label{DMP5.4.0}
\mathcal L f(x) =   \sum_{y\in \Om_N:|y-x|=1}[f(y)-f(x)]
  \end{equation}
  We call $P_{N,x}$ the law of the random walk with $x(0)=x$.

 A function $\la_N(x), x\in \Om_N$, is  harmonic if  $\mathcal L \la_N=0$, namely if  for any $x\in \Om_N$:
  \begin{equation}
 \label{DMP5.4.0}
\la_N(x) = \frac{1}{K_x}  \sum_{y\in \Om_N:|y-x|=1} \la_N(y),\quad
K_x= \big|\{ y\in \Om_N:|y-x|=1\}\big|
  \end{equation}
In the sequel we also need the random walk  $z(t)$, $t\ge 0$ in $\mathbb Z^3$ whose generator is
  \begin{equation}
 \label{DMP5.11}
\mathcal L^0 f(x) =   \sum_{y\in\mathbb Z^3:|y-x|=1} [f(y)-f(x)]
  \end{equation}
We call $P^0_z$ the law of the random walk with $z(0)=z$.
\end{defin}

\bigskip
Observe that  $\la_N(x)\equiv $  constant  is an harmonic function so that Theorem \ref{teorema5.1} includes all DLR measures with an external magnetic field. Thus the interesting point is to find  non constant harmonic functions as proved in \cite{CGP}:

\begin{thm}
 \label{thm5.1}

 For any $\la \ne 0$:

 \begin{itemize}
 \item there is a unique bounded harmonic function $\la^*_N(x)$ such that
    \begin{equation}
 \label{DMP5.5}
\lim_{x_1 \to -\infty}\la^*_N(x) = \la,\quad \lim_{x_1 \to \infty}\la^*_N(x) =-\la
  \end{equation}

 \item $\la^*_N(x)$ is equal to
     \begin{equation}
 \label{DMP5.6}
\la^*_N(x) = \la P_{N,x}\Big[ x(t) \text{\;\;definitively in\;\;$\Om_N^{\rm left}$} \Big]
-\la  P_{N,x}\Big[ x(t) \text{\;\;definitively in\;\;$\Om_N^{\rm right}$} \Big]
  \end{equation}


\end{itemize}
\end{thm}

\medskip

Here is where the condition that the space dimension $d\ge 3$ is essential, in $d<3$ the bounded harmonic functions are constant.  The existence of non constant bounded harmonic functions in our case where $d=3$ also requires a special geometry; an example is $\Om_N$.

\subsection{Stationary magnetization profile}
\label{subsect:4-1}
\medskip
\begin{defin}
\label{mm}We call
	\begin{equation}
	\label{mstar}
	m^*_N(x):= \int \phi_x\,\nu^*_N(d\phi)
	\end{equation}
 where $\nu^*_N$ is the stationary measure of Theorem \ref{teorema5.1} with $\la^*_N$ of Theorem \ref{thm5.1}.  By an abuse of notation we call $m_N(x)$ the stationary magnetization given by \eqref{DMP3.13} with $\nu_N$ the stationary measure in \eqref{DMP3.8} with $\La_{N}$ replaced  by $\La_{N,M}$,  with zero boundary conditions and with
 \begin{equation}
 \label{aaa4.12}
 \la_N(x)=-\frac{\la}{N}x_1
 \end{equation}
Notice that  $\la_N$ is a harmonic function in $\La_{N,M}$ with boundary conditions $\mp \la$ in $x_1 = \pm N$.
\end{defin}

For $\Om_N$ we refer to the figure Fig. \ref{figOm}, where in this case squares represent $\La_{N,M}$, and crosses are the sets $\Sigma^{\pm}$ defined in \eqref{DMP5.7.100}.

\vskip.3cm
We shall prove that  $m^*_N$ and $m_N$ are close in $\La_{N,M}$ by relating  $m_N$ and $m^*_N$ to $\la_N$ and $\la^*_N$ and then using Theorem \ref{thm5.2} below.

%
%

We need  the following  lemma.

\medskip

\begin{lemma}
\label{lemma5.3}
Recalling Definition \ref{defin4.2} we have that for any $\eps>0$: 
   \begin{equation}
 \label{DMP5.10}
\lim_{N\to \infty}\sup_{x:x_1\ge N+R_\eps}P_{N,x}\Big[\text{there is $t$ such that $x(t)\in \Si^+$}\Big] =0
  \end{equation}
and
   \begin{equation}
 \label{DMP5.10bis}
\lim_{N\to \infty}\sup_{x:x_1\le -N-R_\eps}P_{N,x}\Big[\text{there is $t$ such that $x(t)\in \Si^-$}\Big] =0
  \end{equation}
where 
      \begin{equation}
 \label{DMP5.7.100}
R_\eps=M^{2+\eps},\qquad  \Si^{\pm}:=\{x\in \Om_{N}:x_1 = \pm N, |x_2| \le M, |x_3|\le M\}
 \end{equation}

\end{lemma}

\noindent
{\bf Proof.} Fix any $x$ such that $x_1=x_1(0) \ge N+R_\eps$. Call $\tau_+$ the hitting time to $\Si^+$, namely the first time when  $x(t)\in \Si^+$, otherwise $\tau_+=\infty$.  We are going to prove that in law $\tau_+=\tau^0_+$ where $\tau^0_+$ is the hitting time to $\Si^+$ for the random walk
 $z(t)$, $t\ge 0$  defined in Definition \ref{defin4.2} and
  starting from $z(0)=x$.

The main point in the proof is a special realization of the process $x(t)$, already used in \cite{CGP}.  We will realize the process $x(t)$ in the space of the process $z(t)$ and to this end we define
 $\mathcal R$ as
the reflection around the plane $x_1=N-\frac 12$ and define
 \begin{equation}
 \label{DMP5.12}
X(t) =  \begin{cases} z(t) &\text{if $z_1(t)\ge N$}\\\mathcal R(z(t)) & \text{otherwise}
\end{cases}
  \end{equation}
We then have

\begin{itemize}

\item  The law of $X(t)$, $t\le \tau^0_+$, is the same as the law of $x(t), t\le\tau_+$, and $\tau_+$ and $\tau^0_+$ have same law.

\end{itemize}

\noindent
We are thus reduced to study $\tau^0_+$.  Call  $\tau^0_+(y)$, $y\in \Si^+$, the first time when $z(t) =y$, then
   \begin{equation}
 \label{DMP5.13}
P^0_x\Big[ \tau^0_+ < \infty \Big] \le \sum_{y\in \Si^+}
P^0_x\Big[ \tau^0_+(y) < \infty \Big]
  \end{equation}
Let $T=R_\eps^{2-\delta}$, $\delta>0$, then
\begin{equation*}
P^0_x\Big[ \tau^0_+(y) < \infty \Big]\le P^0_x\Big[ \tau^0_+(y) \le T \Big] +P^0_x\Big[ T<\tau^0_+(y) <\infty \Big]
  \end{equation*}
  By the local central limit theorem
  \begin{equation}
 \label{DMP5.14}
P^0_x\Big[ \tau^0_+(y) \le T \Big] \le  e^{- c R_\eps^{\delta}},\quad P^0_x\Big[ T<\tau^0_+(y) <\infty \Big] \le  c'T^{-1/2}
  \end{equation}
Thus
   \begin{equation}
 \label{DMP5.16}
P^0_x\Big[ \tau^0_+ < \infty \Big] \le M^2 \Big(e^{- c R_\eps^{\delta}}
+  c'T^{-1/2}\Big)
  \end{equation}
which vanishes in the limit if we take $\delta<\eps$ small enough  because
  \begin{equation}
 \label{DMP5.17}
 M^2  T^{-1/2}=M^2 R_\eps^{-1+\frac{\delta}2} = M^2 (M^{2+\eps})^{-1+\frac{\delta}2}
 = M^{ \delta-\eps + \eps \frac{\delta}2 }
  \end{equation}
\qed

%
%
%
%
%

By using the above Lemma we have:
\begin{thm}
 \label{thm5.2}
  There is $\eps(N)$, $\dis{\lim_{N\to\infty}\eps(N)=0}$ so that
 \begin{equation}
 \label{eq4.22}
|\la^*_N(x)- \la| \le \eps(N),\quad \forall x_1\le -N,\qquad |\la^*_N(x)+ \la| \le \eps(N),\quad \forall x_1\ge N
  \end{equation}
 \begin{equation}
 \label{eq4.23}
|\la^*_N(x)- \la_N(x)| \le \eps(N),\qquad  -N<x_1<N,
  \end{equation}

\end{thm}

\noindent
{\bf Proof.}   Recalling \eqref{DMP5.6}  and considering $x$ so that $x_1\ge N$, we need to prove that
		\begin{equation}
 \label{eq4.25}
\lim_{N\to\infty} P_{N,x}\Big[ x(t) \text{\;\;definitively in\;\;$\Om_N^{\rm left}$} \Big]=0,\quad
\lim_{N\to\infty} P_{N,x}\Big[ x(t) \text{\;\;definitively in\;\;$\Om_N^{\rm right}$} \Big]=1
  		\end{equation}
Analogous equalities for $x\le -N$, we only prove \eqref{eq4.25}.
		\smallskip
		
Let $\{z(t), t\ge 0\}$ and $\{x (t), t\ge 0\}$ be the random walks defined in Definition  \ref{defin4.2} and assume
$x(0)=z(0)=x \in\Si^+$, thus $x_1(0)=z_1(0)=N$.  Let $\tau_{-N}$ be the first time when $x_1(t)=-N$.

We couple $z(t)$ and $x(t)$ so that till time $\tau_{-N}$ the right jumps  are the same as well as the jumps  to the left when this is possible for $x(t)$.  The transversal jumps are  independent.
Then $x_1(t) \ge z_1(t)$ for all $t\le \tau_{-N}$.  Then from classical theorems  we have that
   \begin{equation}
 \label{DMP5.19}
\lim_{N\to \infty}\inf_{y_1\in[N,N+R_\eps]} P^0_y\Big[\text{$z_1(t)$ reaches $N+R_\eps$ before $-N$}\Big] =1
  \end{equation}
for $R_\eps$  as in \eqref{DMP5.7.100}. Using \eqref{DMP5.10} we then get the second equality in \eqref{eq4.25}. Analogous arguments show the first equality.
We omit the details and give \eqref{eq4.22} for proved.

Calling $\tau_\pm$ the  hitting times to $\Sigma^{\pm}$ and  $\tau=\tau_+ \wedge \tau_-$ we have
	\begin{eqnarray}
\label{DMP3.16b.4}
\la^*_N(x) = E_x \Big[ \la^*_N(x(\tau))\Big],\qquad \forall x\in\La_{N,M}
	\end{eqnarray}
	Thus from \eqref{eq4.22}  we get
	\begin{equation*}
\Big| \la^*_N(x)-[ \la P_x(\tau=\tau_-)-\la P_x(\tau=\tau_+)]\Big|\le \eps(N)
	\end{equation*}
which concludes the proof of \eqref{eq4.23} because the square bracket is equal to
 $\la_N(x)$. \qed

\bigskip

\noindent

\begin{thm}
 \label{thm5.1.1}
 Let $m_N$ and $m^*_N$ as in Definition \ref{mm} then, uniformly in $x\in \La_{N,M}$ such that dist$\big(x,\partial \La_{N,M})> \log N$,
 \begin{equation}
\label{DMP5.7.4}
\lim_{N\to \infty} |m_N(x)-m^*_N(x)|=0
\end{equation}
Furthermore
\begin{equation}
\label{conv}
\lim_{N\to \infty} |m^*_N(x)-\la^*_N(x)|=0
\end{equation}
 \end{thm}

\noindent
{\bf Proof.}
	Proceeding as in Section \ref{sec:3} (see Lemma \ref{lemma3.3} and Theorem \ref{thm3.0.1}) we have  that for all
 $x\in \La_{N,M}$ such that  dist$\big(x,\partial \La_{N,M})> \log N$.
\begin{equation}
	\label{mu3.30a}
|m_N(x) - \la_N(x)|\le c
\Big( \om^{\log N} + \frac{\la}{N}\log N\Big)
		\end{equation}
where $\dis{\om = \frac{6J}{1+6J}}$.
Same holds for $m^*_N(x) -\la^*(x)$ which proves \eqref{conv}. Using \eqref{mu3.30a},  \eqref{conv} and Theorem \ref{thm5.2}, we get \eqref{DMP5.7.4}. \qed

\subsection{Stationary currents}
\label{cur}
We call
\begin{equation}
\label{1}
\hat I_N(x):= \frac 1{(2M+1)^{2}} \sum_{y\in S_x} \int d\nu_N\Big(-NL_{y,y+e_1}\phi_y\Big),\qquad x\in\{-N,..,N-1\}
\end{equation}
 \begin{equation}
\label{1aa}
\hat I^*_N(x):= \frac 1{(2M+1)^{2}} \sum_{y\in S_x} \int d\nu^*_N\Big(-NL_{y,y+e_1}\phi_y\Big),\qquad x\in\{-N,..,N-1\}
\end{equation}
the normalized equilibrium currents through the vertical section 
\begin{equation}
	\label{a4.31}
S_x=\{y\in \La_{N,M}:y_1=x\}
\end{equation}
we will prove that $\hat I_N(x)$ and $\hat I^*_N(x)$ are close to each other for large $N$:
\begin{thm}
\label{teo4}
 $\hat I_N(x)$ and $\hat I^*_N(x)$, $ x\in\{-N,..,N-1\}$ do not depend on $x$
 	\begin{equation}
	\label {aa4.32}
\hat I_N(x)=\mathcal I_N,\quad \hat I^*_N(x)=\mathcal I^*_N, \qquad x\in\{-N,..,N-1\}
	\end{equation}
	$\mathcal I_N$ and $\mathcal I^*_N$ are bounded and
	\begin{equation}
	\label {aa4.33}
\lim_{N\to\infty}\mathcal I_N=\lim_{N\to\infty} \mathcal I^*_N= -\la 	
\end{equation}
\end{thm}

\noindent{\bf Proof.}  Recalling \eqref{aaa4.12} we have
\begin{equation}
\label{a4.34b}
\hat I_N(x)=N[\la_N(x)-\la_N(x+1)]=- \la
	\end{equation}
By the ``conservation of mass"
	\begin{equation*}
\frac d{dt}\int d\nu^*_N \sum_{y\in S_x}\phi_y=0
	\end{equation*}
thus
	\begin{equation*}
 \hat I^*_N(x-1)=\hat I^*_N(x)
	\end{equation*}
	which proves \eqref {aa4.32}.

	To prove that  $\mathcal I^*_N$ is bounded we
take  $x'$ and $x''$  both in $\{-N,..,N-1\}$ with $x'-x''=N$. Since $\hat I^*_N(x)=N[\hat \la^*_N(x)-\hat \la^*_N(x+1)]$ where
	\begin{equation*}
\hat \la^*_N(x)=	\frac 1{(2M+1)^{2}} \sum_{y\in S_x} \la^*_N(y)
	\end{equation*}
 then
	\begin{equation*}
\sum_{x'\le x< x''} \hat I^*_N(x)
=N\mathcal I^*_N= N[\hat \la^*_N(x')-\hat\la^*_N(x'')]
	\end{equation*}
Hence $\mathcal I^*_N$ is bounded because $\la^*_N(x)$ is uniformly bounded by Theorem \ref{DMP5.4.0}.

To prove \eqref{aa4.33} we observe that by \eqref{a4.34b} we only need to prove that the limit of $\mathcal I^*_N$ is $-\la$. Let $\varphi(\xi)$, $-1\le \xi\le 1$ be in $C^1_0(-1,1)$, then
	\begin{eqnarray*}
\frac 1N \sum_{x} \varphi(\frac xN)\hat I^*_N(x)=\frac 1N \sum_{x} \varphi(\frac xN) N[\hat \la^*_N(x)-\hat \la^*_N(x+1)]
	\end{eqnarray*}
Using \eqref{eq4.23} and integrating by parts
\begin{eqnarray*}
\Big|\frac 1N \sum_{x} \varphi(\frac xN)\hat I^*_N(x)-\frac 1N \sum_{x} \varphi(\frac xN)\hat I_N(x)\big| \le
2 \eps(N)\|\varphi'\|_\infty
\end{eqnarray*}
hence $\dis{\lim_{N\to\infty}\mathcal I_N=\lim_{N\to\infty} \mathcal I^*_N}$ which proves \eqref {aa4.33} using \eqref{a4.34b}.\qed

\medskip

\noindent{\bf Remark.} By \eqref{aa4.33}
$\dis{\mathcal I:=
\lim_{N\to\infty}\mathcal I_N=\lim_{N\to\infty} \mathcal I^*_N}$ is equal to $-\la'(r_1)$ where $\la(r)$ is the macroscopic limit of $\la_N(x)$. On the other hand $-\la'(r_1)=-m'(r_1)$ where $m(r)$ is the  limiting magnetization. Hence $\mathcal I =-m'$ in agreement with the Fick law.

\section {Concluding remarks}
\label{sec:6}
As already mentioned the theory developed here can be extended to more general
Ginzburg-Landau stochastic processes provided the temperature is large enough
($\beta$ small).  The extension to large $\beta$ is much more involved and essentially open.  The effect of phase transitions on Fick's law is an interesting open question.

The restriction to Ginzburg-Landau stochastic processes is fundamental in our analysis because for such processes the invariant non-equilibrium measures are known, as proved in \cite{DOP}. The extension to more general systems when $\beta$ is large is essentially open both mathematically and physically.

%
%
\vskip.5cm

\bigskip

\vskip.5cm

\bibliographystyle{amsalpha}

\end{document}